\documentclass[fleqn,usenatbib]{mnras}

\usepackage{newtxtext,newtxmath}

\usepackage[T1]{fontenc}

\DeclareRobustCommand{\VAN}[3]{#2}
\let\VANthebibliography\thebibliography
\def\thebibliography{\DeclareRobustCommand{\VAN}[3]{##3}\VANthebibliography}

\usepackage{graphicx}	
\usepackage{amsmath}	

\title[New Black Hole Spin Values for Sgr $\rm{A^*}$]{New Black Hole Spin Values for Sagittarius $\rm{A^*}$ Obtained with the Outflow Method}

\author[Daly et al.]{Ruth A. Daly,$^{1}$\thanks{E-mail: rdaly@psu.edu}
Megan Donahue,$^{2}$
Christopher P. O'Dea$^{
3}$, 
Biny Sebastian$^{3}$, Daryl Haggard$^{4,5,6}$, 
and Anan Lu$^{4,5}$
\\
$^{1}$Department of Physics, Penn State University, Berks Campus, Reading, Pennsylvania, 19610-6009, USA\\
$^{2}$Michigan State University, Physics \& Astronomy Department, East Lansing, MI, USA\\
$^{3}$Department of Physics \& Astronomy, University of Manitoba, 30A Sifton Rd., Winnipeg, MB R3T 2N2, Canada\\
$^{4}$Department of Physics, McGill University, 3600 University Street, Montréal, QC H3A 2T8, Canada\\
$^{5}$McGill Space Institute, McGill University, 3550 University Street, Montréal, QC H3A 2A7, Canada\\
$^6$CIFAR Azrieli Global Scholar, Gravity \& the Extreme Universe Program, Canadian Institute for Advanced Research, 661 University Avenue, \\~~~Suite 505, 
Toronto, ON M5G 1M1, Canada\\
}

\date{Accepted XXX. Received YYY; in original form ZZZ}

\pubyear{2015}

\begin{document}
\label{firstpage}
\pagerange{\pageref{firstpage}--\pageref{lastpage}}
\maketitle

\begin{abstract}
Six archival Chandra observations are matched with eight sets of 
radio data and studied in the context of the outflow method to measure and study the 
spin properties of $\rm{Sgr ~A^*}$. 
Three radio and X-ray data sets obtained simultaneously, or 
partially simultaneously, are identified as 
preferred for the purpose of measuring the spin properties  of $\rm{Sgr ~A^*}$. Similar results are  
obtained with other data sets. Results 
obtained with the preferred data sets are combined and 
indicate a weighted mean value of the spin 
function of $\rm{F} = 0.62 \pm 0.10$ and dimensionless 
spin angular momentum of
$\rm{a_*} = 0.90 \pm 0.06$.  
The spin function translates into measurements of 
the black hole rotational mass, 
$\rm{M_{rot}}$, irreducible mass, $\rm{M_{irr}}$, and spin mass-energy available 
for extraction, $\rm{M_{spin}}$, relative to the total black hole dynamical mass, $\rm{M_{dyn}}$. 
Weighted mean values of  
$\rm{{(M_{rot}/M_{dyn})} = (0.53 \pm 0.06)}$, $\rm{({M_{irr}/M_{dyn})} = (0.85 \pm 0.04)}$, $\rm{({M_{spin}/M_{dyn})} = (0.15 \pm 0.04)}$,  
 $\rm{{M_{rot}} = (2.2 \pm 0.3) \times 10^6 ~M_{\odot}}$, ~$\rm{{M_{irr}} = (3.5 \pm 0.2) \times 10^6 ~M_{\odot}}$, and $\rm{{M_{spin}} = (6.2 \pm 1.6) \times 10^5 ~M_{\odot}}$ are obtained; of course 
 $\rm{{(M_{rot}/M_{irr})} = (0.62 \pm 0.10)}$ since 
 $\rm{{(M_{rot}/M_{irr})} = F}$. 
Values obtained for $\rm{Sgr ~A^*}$ 
are compared with those obtained for M87 based on the published spin function which 
indicate that M87 carries substantially more rotational energy and 
spin mass-energy relative to the total (i.e., dynamical) black hole mass, the 
irreducible black hole mass, and in absolute terms.  
\end{abstract}




\section{Introduction}
Sagittarius A$^*$ (Sgr A$^*$) is the supermassive black hole that resides at the center of the Milky Way Galaxy. The total black hole mass, also referred to as the dynamical mass, 
$\rm{M_{dyn}}$, 
is known to high accuracy (e.g Gravity Collaboration 2019). 
Both the black hole irreducible mass, $\rm{M_{irr}}$, and the rotational mass, 
$\rm{M_{rot}}$ contribute to the dynamical black hole mass: 
$\rm{M_{dyn}^2 = M_{irr}^2 + M_{rot}^2}$ (e.g., Misner, Thorne, \& Wheeler 1973). The spin mass-energy 
available for extraction is $\rm{M_{spin} = M_{dyn} - M_{irr}}$ (e.g., Thorne et al. 1986); this is the spin mass-energy that is available to, and could in principle, power outflows and jets, for example.   
Extraction of spin mass-energy from a black hole can have a significant impact 
on the black hole environment, and can decrease the black hole dynamical mass (e.g., Penrose 1969; Christodoulos 1970; Penrose \& Floyd 1971). 
In addition, the gravitational impact of a spinning black hole on bodies and 
material in the immediate environment of the black hole is significantly  
different from that of a non-spinning black hole. For these reasons, it is 
interesting and important to empirically determine the spin properties of 
Sgr A$^*$. At present, most studies of the spin properties 
of Sgr A$^*$ are highly model dependent, and it 
appears that the community has not reached a consensus 
regarding the spin properties of this source (see the discussion in section 4). 

Ratios of the rotational mass, irreducible mass, and spin mass-energy available for extraction relative to the  dynamical mass of the black hole can be determined if the spin function or dimensionless spin angular momentum of the black hole is known (e.g., Christodoulou et al; Christodoulos 1970;
Misner, Thorne, \& Wheeler 1973; Rees 1984; MacDonald \& Thorne 1982; Thorne et al. 1986). These relationships 
have been expanded and applied to study the spin properties of various samples of sources (Daly 2009,  2022). 
For example, Daly (2022) showed that the spin function, $\rm{F = (M_{rot}/M_{irr})}$, and the ratios  
$\rm{(M_{rot}/M_{dyn})}$, $\rm{(M_{irr}/M_{dyn})}$, 
$\rm{(M_{spin}/M_{dyn})}$, and the spin mass-energy 
available for extaction relative to the maximum possible 
value, $\rm{(E_{spin}/E_{spin,max})}$, can be obtained 
directed from F. (Note that the nomenclature has been simplified here to refer to F as the spin function, whereas in Daly (2019, 2022) $\rm{F^2}$ was referred to as the ``spin function.") The ratios listed above can be combined with 
the empirically determined dynamical mass to obtain values for $\rm{M_{rot}}$, $\rm{M_{irr}}$, 
$\rm{M_{spin}}$ in units of solar masses. 

Here, X-ray and radio data are considered in the context of 
the outflow method and are applied to empirically determine the spin function, F, of Sgr A$^*$, and 
all of the quantities that can be determined from the spin function. 
The data are described in section 2. The outflow method is described in section 3. 
The results are presented in section 4. A discussion 
of the results follows in section 5. In section 5, quantities obtained 
for Sgr A$^*$ are compared with those obtained for 
M87. Results for M87 follow from the spin function published by Daly (2019) obtained with 
the outflow method. 

\section{Data Selection and Analysis}
\subsection{Selection of Simultaneous and Contemporaneous Data}
Six archival Chandra X-ray observations of Sgr $\rm{A^*}$ that are simultaneous with, or partially simultaneous with, four radio data sets from Capellupo et al. (2017) and contemporaneous with 
four individual radio observations from Bower et al. (2015) 
are identified. 
Table 1 lists the Chandra observation identifications (IDs) and dates, and the 
intrinsic (2-10) keV flux densities, as 
described in section 2.2. 
The radio data sets are described in section 2.3. The simultaneously 
obtained (or partially simultaneously obtained) radio data sets from 
Capellupo et al. (2017) are summarized in Table 2. Radio data from 
Bower et al. (2015) obtained contemporaneously with X-ray data are 
summarized in Table 3. 

\subsection{The (2-10) keV X-ray Luminosities}

Chandra archival X-ray data are analyzed to obtain the intrinsic (2-10) keV X-ray flux density 
for the six different observations of Sgr $\rm{A^*}$. 
The X-ray data are downloaded from the Chandra archive and reprocessed using CIAO 4.15 and CALDB 4.10.2. The CIAO (Fruscione et al. 2006) routine srcflux was used to generate source and background spectra, and corresponding aperture-corrected fluxes. A relatively compact region was chosen ($r=1^{\prime\prime}$), centered on RA, Dec (J2000) of $17^h~45^m~40.125^s$, $-29^d~00'~28.24"$ to avoid a nearby bright X-ray source. An annular background nearly concentric with Sgr A$^*$ with inner/outer radii of 4.92-6.74 arcseconds was used to extract a background spectrum. In all of the observations, the background is a negligible contribution to the total count rate ($<10-12\%$ in each case.) Observation-specific response (rmf) and ancillary response (arf) matrices were generated for each spectrum.

We used XSPEC 12.12.1 (Arnaud 1996) to analyze the X-ray spectra. 
We simultaneously fit a single spectral shape of an absorbed power-law spectrum to the X-ray data between 2-7 keV, allowing the normalizations (or, equivalently, the inferred 2-10 keV flux) to be different for different datasets. These choices allowed for a self-consistent estimate of the power-law slope and absorbing column, which were not well-constrained with individual spectra. We obtained a common photon index of $2.0\pm 0.3$  and an X-ray column density of $1.1 \pm 0.1 \times 10^{23}~\rm{cm}^{-2}$. These quantities are consistent with the expected photon index of 2 and approximate column density of Sgr A$^*$.
We used XSPEC mcmc-chains to derive the 67\% uncertainty ranges for the corresponding intrinsic (2-10) keV fluxes, reported in Table 1. 
These observations are independently studied by 
Bagnaoff et al. 2001; Nowak et al. 2012; Wang et al. 
2013; and Neilsen et al 2013, for example.

\begin{figure}
\includegraphics[width=0.45\textwidth]{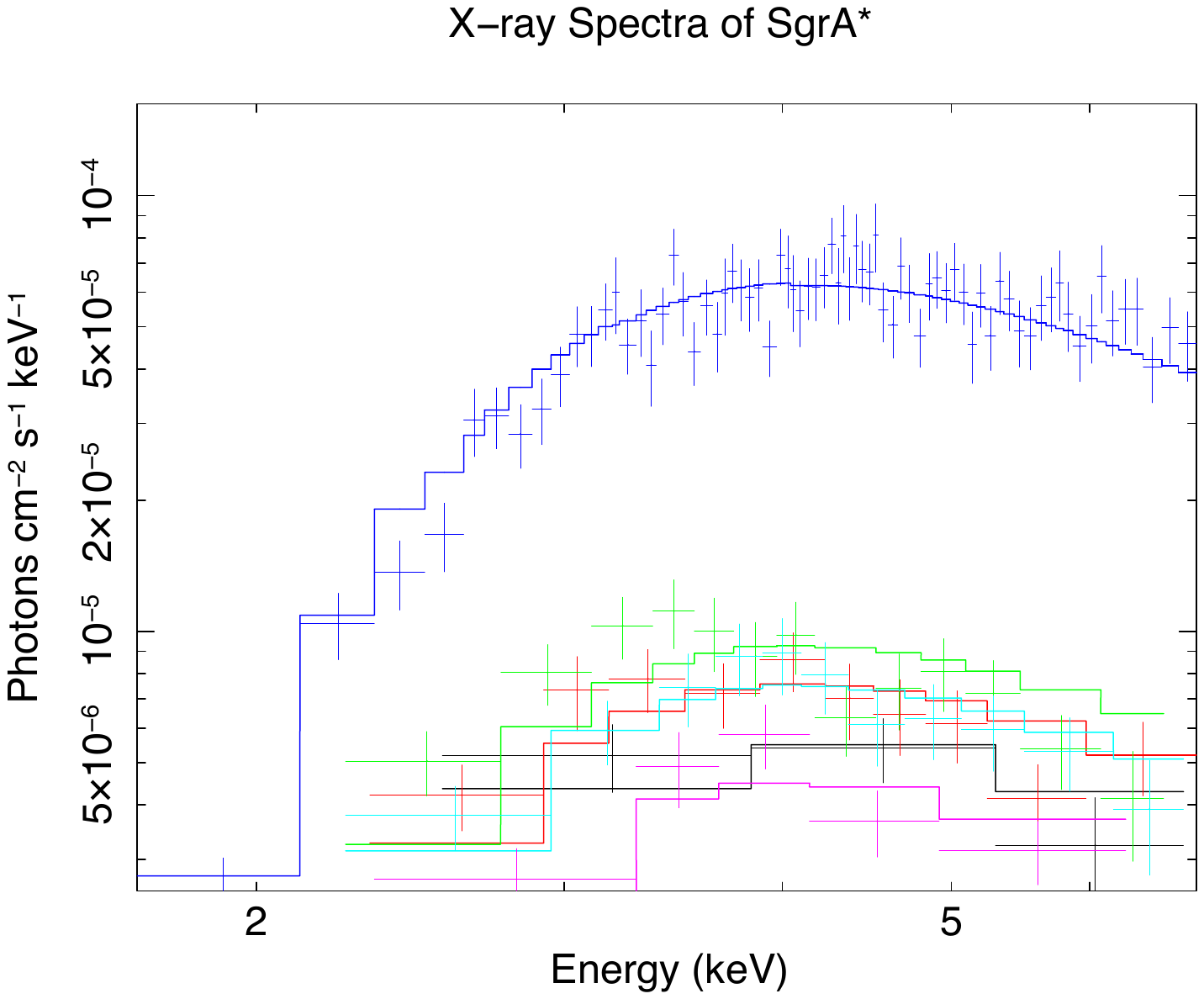}
\caption{The Chandra X-ray spectra (the data normalized by the response and exposure time) and the corresponding best fit models for the six observations of Sgr A$^*$ studied here. These are binned to a minimum of $5\sigma$ per energy bin for visualization. The brightest spectrum is from 14-Sept-2013, ObsID 15043; the other spectra shown are more typical for this source. }
\end{figure}

The X-ray flux density varies and flickers at a low level during 3 of the X-ray observations. By far, the most dramatic outburst occurred during observation ID 15043 
(Haggard et al. 2019). 
Indeed, considering the 70 Chandra observations of Sgr A$^*$ obtained between 
2000 and 2014, this outburst is the brightest X-ray event recorded by the Chandra X-ray Observatory, and it is a true outlier (e.g., Neilsen et al. 2013). 
For example,  only two other events, one about half as bright and one about a third as bright, were observed over this time period (e.g., Nowak et al. 2012; Haggard et al. 2019). 
However, as explained in section 3 and shown in section 4, 
the inclusion or the exclusion of the X-ray 
flaring event and its corresponding radio counterpart only marginally impact the results.

\subsection{The 5 GHz Radio Luminosities}

The application of the outflow method to measure the spin properties of black holes for sources such as Sgr A$^*$ requires the 5 GHz radio luminosity, as explained in section 3. VLA radio  
observations obtained simultaneously with 
Chandra X-ray observations were 
reported by 
Capellupo et al. (2017) at (8-10) GHz; each epoch 
of radio observation includes numerous observations per day (see Table 2). 
The Capellupo et al. (2017) radio data sets listed in Table 2 are labelled  
C1 - C4.
The mean value and standard deviation of the (8-10) GHz flux from 
Capellupo et al. (2017) are 
listed in Table 2 along with the number, N, 
of individual VLA observations available for each run and applied to obtain the mean radio flux density for that data set. 

To shift the (8 - 10) GHz VLA data of Capellupo et al. (2017) to 5 GHz the (5 to 9) GHz radio spectral index is required. 
The data published by Bower et al. (2015) included both 5.4 and 8.9 GHz observations of Sgr A$^*$ obtained on the same day for 11 different days. These were used to determine  the radio spectral index between these two frequencies. 
The weighted mean value of the eleven 5.4 to 8.9 GHz spectral indices indicate a mean value of $\alpha$ of  $0.16 \pm 0.03$, where the flux density 
is written as $f \propto v^{\alpha}$(i.e., a positive value of $\alpha$ indicates that the flux density is increasing with frequency). 
This value is consistent with that reported by 
Melia \& Falcke (2001). The uncertainty in using this spectral index to 
scale the radio flux density  
of the Capellupo et al. (2017) or Bower et al. (2015) data to 
5 GHz was added in quadrature with the rms 
dispersion of the radio flux density to obtain the one-sigma uncertainty of 
the 5 GHz flux density for both the 
(8 - 10) GHz radio data obtained by Capellupo et al. (2017) and the 
5.4 GHz radio data reported by Bower et al. (2015), described below.  
 
Four VLA 5.4 GHz radio observations 
reported by Bower et al. (2015) were obtained within a few days 
of Chandra X-ray observations (described in section 2.2). The relevant information for these observations are listed in Table 3. 
The Bower et al. (2015) radio data sets listed in Table 3 are labelled  
B1 - B4. 
The B2 and B3 radio observations from Bower et al. (2015) were obtained within a few days of the C2 and C3, so 2 of the X-ray observations were applied twice, 
in combination with both B2 and C2, and with B3 and C3.

The 5 GHz flux density is converted to a 5 GHz luminosity by multiplying the flux density by the observed frequency of 5 GHz and adopting a value of $8.178 \pm 0.013$ kpc to the source Sgr $\rm{A^*}$; the 
intermediate value presented in Table 1 of the Gravity Collaboration (2019) listed 
as the "noise model fit" was applied. The mass of Sgr A\* adopted is from the same fit and is 
$(4.152 \pm 0.014) \times 10^6 M_{\odot}$.   

\begin{table}								
\begin{minipage}{80mm}							
\caption{Intrinsic (2-10) keV Chandra Flux Densities}
\label{Table1}								
\begin{tabular}{llc}   
\hline\hline  
(1)	&	(2)	&	(3)\\
Chandra&Chandra	&$	f ~\rm{(2-10)~ keV}	$\\
   Obs ID	&	Date	&$	10^{-13}\rm{(erg ~s^{-1} ~cm^{-2})}$\\
\hline
$14703	$&$	20130604	$&$	2.38\pm	0.45$\\
$15041	$&$	20130727	$&$	3.07 \pm	0.23$\\
$15042	$&$	20130811$&$	3.74\pm	0.25$\\
$15043	$&	 20130914 	&$	26.02	\pm	1.27$\\
$15045	$&$20131028$&$	3.23	\pm	0.26	$\\
$16213	$&$20140428$&$	1.80	\pm	0.21	$\\
\hline
\hline
\end{tabular}									
\end{minipage}									
\end{table}

\begin{table}								
\begin{minipage}{80mm}							
\caption{(8-10) GHz VLA Flux Densities Obtained Simultaneously with Chandra X-ray  Data (discussed by Capellupo et al. 2017)}
\label{Table2}								
\begin{tabular}{llllll}   
\hline\hline  
(1)&(2)&(3)&(4)&(5)&(6)\\	
&	Chandra & Chandra	&	Radio	&N&$f$~\footnote{Mean value and standard deviation of the mean obtained with N observations.}	\\
&   ObsID&ID Date&ID Date&&(Jy)  \\
\hline
C1~\footnote{The Capellupo et al. (2017) VLA radio datasets applied here are labelled C1 through C4.}	& 15041 & 20130727	&$	20130727	$&320&$0.923 \pm 0.055$ \\
C2	& 15042 & 20130811	&$	20130812	$&304&$0.770 \pm 0.049$ \\
C3	& 15043 & 20130914~\footnote{The Chandra X-ray data for this epoch includes a bright X-ray flare that increases the total X-ray luminosity by about a factor of ten; this X-ray data 
is also used in conjunction with both the B3 radio data.}	&$	20130913	$&312&$0.975 \pm 0.075$ \\
C4~\footnote{This data set is displayed in Fig. 1 of Capellupo et al. (2017) offset 
by a factor of 1.32.}	& 16213 & 20140428	&$	20140428	$&304&$0.743 \pm 0.011$\\
\hline
\hline
\end{tabular}									
\end{minipage}									
\end{table}

\begin{table}								
\begin{minipage}{80mm}							
\caption{Contemporaneous X-ray and Radio Observations of $\rm{Sgr A}^*$  }						
\label{Table3}								
\begin{tabular}{llllll}   
\hline\hline  
(1)&(2)&(3)&(4)&(5)&(6)\\	
Radio&	Chandra & Chandra	&	Radio	&	$\nu$&N	\\
Data~\footnote{The Bower et al. (2015) radio datasets are labelled B1 through B4.}&   ObsID &  Date&Date&  (GHz)  \\
\hline
B1	& 14703 & 20130604	&$	20130609	$&  5.4&1 \\
B2	& 15042 & 20130811	&	20130808	&5.4&1  \\
B3	& 15043 & 20130914~\footnote{The Chandra X-ray data for this epoch includes a bright X-ray flare that increases the total X-ray luminosity by about a factor of ten; this X-ray data 
is also used in conjunction with  the C3 radio data.}	&$	20130918	$&  5.4&1\\
B4	& 15045 & 	20131028	&$	20131026	$& 5.4&1 \\
\hline
\hline
\end{tabular}									
\end{minipage}									
\end{table}

Results for each of the radio data sets listed in Tables 2 and 3 are obtained and 
presented in Section 4. When results from data sets are combined, the 
B2 and B3 results are not included since the C2 and C3 data sets, 
which include numerous individual radio observations and which 
are obtained partially simultaneously with X-ray data, are preferred to the 
single B2 and B3 radio observations which are obtained contemporaneously 
with the X-ray data.  

\section{Method}
 
The outflow method of measuring the spin properties of a black hole 
is based on the premise that black hole spin angular momentum and energy powers a 
collimated outflow or dual collimated outflows, either in part or in full, 
that emanate from a black hole system for certain types of systems (Daly 2016, 2019). 
The ''black hole system" includes the 
black hole, the accretion disk (which refers to gaseous material in the vicintiy of the black hole), and the collimated outflow. 
The outflow method was applied to $\rm{Sgr~ A^*}$ by Daly (2019) who 
report a dimensionless spin angular momentum of $\rm{a_* = 0.93 \pm 0.15}$ 
(see Tables 1 and 3 of that paper; note that the dimensionless 
spin angular momentum was referred to with the symbol $\rm{j}$ in that work 
while here it is referred to with the symbol $\rm{a_*}$). 
The dimensionless black hole spin parameter 
$\rm{a_* \equiv J c/(GM^2)}$ where $\rm{J}$ is the spin angular momentum of the 
black hole, $\rm{M}$ is the total black hole mass (also referred to as 
the dynamical black hole mass, since this is the mass that will be 
measured using the local dynamics of the black hole region or any 
other astronomical observation), $c$ is the speed of light, and $G$ 
is Newton's constant (e.g., Misner, Thorne, \& Wheeler 1973). 
In this paper, 
the same method is applied to new and larger data sets to obtain updated  
black hole spin measurements  of $\rm{Sgr~ A^*}$. 

The motivation for the method 
and the derivation of the primary equations that describes the method are summarized in section 1.1 of Daly (2019). The outflow method is 
motivated by the functional form of empirically determined 
relationships, and does not rely upon any specific 
jet-powered outflow model or any specific accretion disk model. 
The black hole spin is parameterized by the spin function 
F where $\rm{F = a_*({1+ [1-a_*^{2}}]^{0.5})^{-1}}$ and is 
empirically determined by applying eq. (2) of Daly (2019): 
\begin{equation}
\rm{F^2 = (L_{dKE}/(g_j L_{Edd})) ~(L_{bol}/(g_{bol}L_{Edd}))^{-A}},  
\end{equation}
where $\rm{L_{dKE}}$ is the luminosity in directed kinetic energy 
carried by the collimated outflow (also referred to as the 
beam power, $\rm{L_j}$), $\rm{L_{bol}}$ is the 
bolometric disk luminosity of the AGN accretion disk, 
$\rm{L_{Edd}}$ is the Eddington luminosity  obtained from 
the dynamical black hole mass, and $\rm{g_j}$ and $\rm{g_{bol}}$ are normalization 
factors for the beam power and bolometric luminosity, respectively. 
The parameter $A$ is obtained as described in section 
3.2 of Daly (2019) and by Daly et al. (2018), and for sources 
such as $\rm{Sgr~ A^*}$ and M87 is $0.41 \pm 0.04$ (see line 3 of Table 2 
of Daly et al. 2018). This follows because $\rm{Sgr~ A^*}$ and M87 
are included in the 
sample of AGN studied by Merloni et al. (2003) who identified 
the fundamental plane of black hole activity for supermassive and 
stellar mass black holes (see also Falcke et al. 2004). 
That is, since $\rm{Sgr~ A^*}$ is included in the Merloni et al. (2003) 
sample, the properties of this data set and fits to this data set 
are used to study $\rm{Sgr~ A^*}$ in the context of the outflow method.  
The fundamental 
plane of black hole activity is a relationship between the 
radio luminosity of the jetted outflow (Merloni et al. 2003 used 
the 5 GHz rest frame luminosity), the X-ray luminosity of the 
source (Merloni et al. 2003 used the (2-10) keV luminosity), and the 
black hole mass. As discussed, for example, by Merloni \& Heinz (2007) 
the radio luminosity is most likely related to the outflow beam 
power, $\rm{L_{dKE}}$, the X-ray luminosity is most likely related to the 
bolometric disk luminosity, $\rm{L_{bol}}$, of the AGN 
accretion disk, 
and of course the black hole mass 
is related to the Eddington luminosity, $\rm{L_{Edd}}$. 

Daly et al. (2018) showed that the beam power 
(i.e., luminosity in directed kinetic energy) 
for radio sources 
that lie on the fundamental plane of black hole activity can be 
empirically determined from the functional form of that plane and the radio luminosity 
used to define the plane (e.g., see eq. 4 and the values 
of C and D listed in Table 1 of Daly et al. 2018). This is 
accomplished by mapping the fundamental plane of 
black hole activity to the fundamental line of black hole activity, 
where the fundamental line has the 
functional form $\rm{Log(L_{dKE}/L_{Edd}) = A~Log(L_{bol}/L_{Edd}) +B}$. 
That is, identifying the fundamental plane 
of black hole activity as the empirical manifestation 
of the fundamental line of black hole activity, 
Daly et al. (2018) found that mapping the 
fundamental 
plane (of black hole activity)  
to the fundamental line (of black hole activity)  
provides a method to 
empirically determine the outflow beam power
(i.e., luminosity in directed kinetic energy) of each 
source that lies on the fundamental plane. 
This method does not require the use of 
a detailed physical model for each or any of the sources in the 
sample used to define the fundamental plane for that sample. 
This method of obtaining the outflow beam power 
is referred to as the ``fundamental line mapping method" (FLMM). 
This is how the beam powers (i.e., $\rm{L_{dKE}}$) 
presented here are obtained.  
The well-known relationships between the 
(2-10) keV X-ray luminosity and bolometric luminosity (e.g., Ho 2009; Daly et al. 2018), and between 
the black hole mass and Eddington luminosity were used 
to obtain those quantities. The fact that a well-defined plane such as 
the fundamental plane is obtained using the (2-10) keV luminosity of each source 
indicates that the same conversion factor 
should be applied to each of the sources in the sample to convert the X-ray luminosity 
to the bolometric luminosity, which is the intrinsic fundamental physical variable.  
Similarly, the same conversion method should be 
applied to map the radio luminosity to the beam power (i.e., 
luminosity in directed kinetic energy) for each of the sources in the 
sample, which is the intrinsic fundamental physical variable. 

Mapping the fundamental plane of black hole activity to the 
fundamental line of black hole activity for several different 
fundamental plane samples, Daly et al. (2018) found that the 
dispersion of the fundamental line was significantly smaller than 
that of the fundamental plane.
This indicated that the intrinsic 
relationship is described by the fundamental line, and that the 
fundamental plane is the empirical manifestation of the true 
underlying relationship described by the fundamental line. 
Given the known uncertainties of the bolometric luminosity and 
Eddington luminosity, the very small dispersion of the fundamental 
line indicated that the uncertainty of the beam power was significantly smaller than that obtained by blindly 
propagating uncertainties that enter through the mapping of the fundamental plane to the fundamental line (likely because in reality 
it is the plane that results from the line rather than the other way around). 
To determine the uncertainty of the beam power obtained with the FLMM, 
Daly (2019) combined the dispersion of the fundamental line for the 
Merloni et al (2003) sample with that obtained by Daly (2016) for sample of 
powerful classic double (``FRII") radio sources for which all quantities were obtained 
with completely independent methods, along with the 
known uncertainty of the beam poower for 
each FRII source, to obtain the uncertainty 
of the beam power for each fundamental 
plane source in the Merloni et al. (2003) sample. This indicated an uncertainty of 
$\rm{\delta{Log(L_{dKE}))}}= 0.24$ for beam 
powers obtained with the fundamental line 
mapping (FLMM) method for 
sources in the Merloni et al. 
(2003) sample including  Sgr A$^*$ and M87,  
as described in detail in section 2 
of Daly (2019). This uncertainty is applied here. 

The parameters $\rm{g_{bol}}$ and $\rm{g_j}$ introduced 
by Daly (2016) for a different category of source were studied 
for four types of sources by Daly et al. (2018) who determined 
that $\rm{g_{bol} = 1}$ and $\rm{g_j = 0.1}$; and these values were used 
by Daly (2019) and are adopted here. These are close to the theoretically expected values, as discussed in section 4 of Daly (2019).

\section{Results}

Black hole spin characteristics for $\rm{Sgr~A^*}$ are summarized in Tables 4 - 7. The values 
of the spin function, F, and the dimensionless spin angular momentum,  
$\rm{a_*}$, presented here can be compared with those reported by Daly (2019), who applied the outflow method and obtained a value of 
$\rm{F} = 0.68 \pm 0.30$ indicating a value of $a_* = 0.93 \pm 0.15$ 
for Sgr A* 
(see Tables 1 and 3 of that work). 
The values obtained here and listed in Table 4 are consistent with the previously reported value, 
and have smaller uncertainties. 
The spin function, F, and dimensionless spin 
angular momentum, $\rm{a_*}$, for M87 obtained and reported by Daly (2019) 
are also included in Table 4, and results obtained 
with that spin function are included in Table 5. 

Traditionally, the 
relationships between $\rm{M_{rot}}$, $\rm{M_{spin}}$, $\rm{M_{irr}}$, 
and $\rm{M_{dyn}}$ have been written in terms of 
the dimensionless black hole spin angular momentum, $\rm{a_*}$. However, the form of these equations 
does not allow for values of $\rm{a_*}$ greater than one. Measurement uncertainties are expected to 
lead to values of $\rm{a_*}$ greater than one, especially for 
highly spinning black holes. To circumvent this issue, Daly (2022) recast the relationships 
between $\rm{M_{rot}}$, $\rm{M_{spin}}$, $\rm{M_{irr}}$, 
and $\rm{M_{dyn}}$ 
in terms of the spin function $\rm{F}$, which allows 
for values of $\rm{F}$ greater than one; note that 
a maximally spinning black hole corresponds 
to a value of F of one. This is valuable in understanding the theoretical implications of  empirically determined values of F, which may be  greater one due to measurement uncertainties that enter into the 
empirically determined quantities used to measure F (Daly 2022). 

Six of the values of F reported here $\rm{Sgr~A^*}$ are obtained with independent Chandra and radio data sets. 
Results obtained with the Capellupo et al. (2017) radio data and Chandra archival data are  
the most reliable because the radio luminosity is based on an average obtained over the course of 
the day and because the radio and X-ray data were obtained simultaneously or 
partially simultaneously. Radio data  obtained at (8 - 10) GHz was converted to 5 GHz since the mapping from radio luminosity to the luminosity in directed kinetic energy (i.e., the beam power) is based on the 5 GHz radio luminosity, 
as described in sections 2.3 and 3. 
Results obtained with the 
flaring X-ray data (C3 and/or B3)
are not representative of the 
typical behavior of $\rm{Sgr~A^*}$, and thus should 
be discounted as described in section 2.3.  
Therefore, the "preferred" results for 
$\rm{Sgr~A^*}$ are those obtained with the three 
non-flaring X-ray observations obtained simultaneous with the 
Capellupo et al. (2017) data sets, 
C1, C2, and C4, referred 
to as Set I in Tables 6 and 7. For 
completeness, results obtained with other combinations 
of data sets are also included in Tables 6 and 7. 

Values of the spin function, F, indicate the dimensionless spin angular momentum, $\rm{a_*}$, and  the 
rotational mass-energy, $\rm{M_{rot}}$, the irreducible black hole mass, $\rm{M_{irr}}$, and the spin mass-energy available for extraction, $\rm{M_{spin}}$, relative to the dynamical black hole mass, $\rm{M_{dyn}}$, as discussed in detail 
by Daly (2022); see  equations 
(9) - (15) of that work. Of course, these can be combined with 
the measured black hole mass to obtain 
$\rm{M_{rot}}$, $\rm{M_{irr}}$, and $\rm{M_{spin}}$ in 
physical units. 
These are important parameters that quantify 
the different components that contribute to the total 
dynamical black hole mass of $\rm{Sgr~A^*}$, and the 
mass-energy that can be extracted. 
The irreducible black hole mass cannot be decreased (except by Hawking radiation, which is not expected to be effective except on extraordinarily long timescales). 
The rotational mass indicates the mass-energy 
contribution of rotation to the total dynamical mass, 
where $\rm{M_{dyn}^2} = \rm{M_{irr}}^2 + \rm{M_{rot}}^2$ 
(e.g., Misner, Thorne, \& Wheeler 1973).  
Only part of the rotational mass-energy is available 
to be converted into a useful form that could power an outflow, 
$\rm{M_{spin}} = \rm{M_{dyn}} - \rm{M_{irr}}$  
(e.g., Rees 1984; Thorne et al.1986). 
Other interesting parameters to study are the 
fraction of the rotational mass that is available to be extracted and converted into a useful form,  
$(\rm{M_{spin}/M_{rot}})$, the spin mass-energy relative to the 
irreducible mass, $(\rm{M_{spin}/M_{irr}})$, and the spin energy 
available to be extracted relative to that expected for a maximally 
spinning black hole, 
$\rm({E_{spin}/E_{spin,max}})$. 
Values of these parameters 
are listed in Table 5 for each of the eight radio data sets considered for Sgr A$^*$; values for M87 are also listed, which 
are obtained with the spin function 
listed in Table 4. 

Relative and absolute values of these mass-energy components for $\rm{Sgr~A^*}$ obtained with 
six Chandra data sets 
associated with eight radio data sets are listed in Table 5; 
results for M87 are also included in that table with the value of $F$ and 
$a_*$ for that source obtained from Daly (2019). For Sgr A*, 
the rotational mass is about half the value 
of the total dynamical mass, as is evident from column (2) 
of Table 5. The "preferred" value obtained here, 
described above and 
referred to as Set I, is listed in the first row of Tables 6 has a value 
of $(\rm{M_{rot}}/\rm{M_{dyn}}) = (0.53 \pm 0.06)$. 
Combining this with the dynamical black hole mass 
of $(4.152 \pm 0.014) \times 10^6 ~\rm{M_{\odot}}$ indicates that 
the rotational mass of $\rm{Sgr~A^*}$ is 
$\rm{M_{rot}} = (2.2 \pm 0.3) \times 10^6~\rm{M_{\odot}}$. 
Preferred values of other mass-energy characteristics 
for $\rm{Sgr~A^*}$ are listed in the first row  of 
Tables 6 and 7 and are: 
$(\rm{M_{irr}}/\rm{M_{dyn}}) = (0.85 \pm 0.04)$ and 
$\rm{M_{irr}} = (3.5 \pm 0.2) \times 10^6 ~\rm{M_{\odot}}$;
$\rm({M_{spin}}/\rm{M_{dyn}}) = (0.15 \pm 0.04)$ and 
$\rm{M_{spin}} = (6.2 \pm 1.6) \times 10^5 ~\rm{M_{\odot}}$;
$\rm({M_{spin}}/\rm{M_{rot}}) = (0.29 \pm 0.04)$;
$\rm({M_{spin}}/\rm{M_{irr}}) = (0.18 \pm 0.05)$;
$\rm({E_{spin}}/\rm{E_{spin,max}}) = (0.42 \pm 0.13)$; 
$\rm({M_{rot}}/\rm{M_{irr}}) = F = (0.62 \pm 0.10)$; 
and $a_* = (0.90 \pm 0.06)$.

The highly non-linear relationship between the dimensionless spin angular momentum, $a_*$, 
and other parameters that characterize the spin properties 
of black holes suggests that it is beneficial to 
use alternative parameters to study the spin 
properties of black holes, such as those considered here. 

The results obtained here regarding the 
dimensionless spin angular momentum are consistent with those 
reported by Huang et al. (2009),  
Mościbrodzka et al. (2009), Eckart et al. (2018),  
Gravity Collaboration (2019), and 
Event Horizon Telescope Collaboration (2022), 
all of which are 
obtained with methods independent of those considered here. 

The review of Sgr A* black hole properties by Eckart et al. (2018) presents estimates of the dynamical mass, dimensionless spin angular momentum, charge, and orientation.  Their Table 3 lists ten representative values of the dimensionless spin parameter $a_*$ determined using several different techniques. The estimates cover the range from 0 to 1, though Eckart et al. (2018) suggest the spin is likely to be between 0.5 and 0.92.  Additional estimates in the literature are
Fragione \& Loeb (2020) $a_* < 0.1$; 
Broderick et al. (2016) $a_* \sim 0.1$; 
Huang et al. (2009) $a_* < 0.9$; 
Mościbrodzka et al. (2009) $a_* \sim 0.9$; 
Shcherbakov et al.  (2012) 
considered models with spin values of 
$0, 0.5, 0.7, 0.9, \& ~
0.98$ and found the ``best-bet" 
model has $a_* \sim 0.5$. 
Running a series of accretion disk 
models and considering  specific values of $a_*$ such as 
${0, 0.5, 0.7, \& ~0.94}$
in different combinations 
with different models and 
followed by a 
comparison of simulation outputs with 
Event Horizon Observations (Event Horizon Collaboration 2022) the 
observations favors the high-spin models 
considered over the low-spin models considered. Thus, previous estimates are consistent with a rotating black hole, though there is no agreement on the value of the dimensionless spin parameter.

\section{Discussion}

Results obtained for Sgr A* in the context 
of the outflow method and presented above 
are set in the context of our broader knowledge of Sgr A* in this section.  

1). There is substantial evidence for a collimated outflow, also referred to as a jet, from Sgr A* 
(e.g., Falcke \& Markoff 2000; 
Li et al. 2013; 
Brinkerink et al. 2015; 
Zhu et al. 2019; 
Yusef-Zadeh et al. 2020; Brinkerink et al. 2021). 
The details of jet production 
close to 
Sgr A* have been discussed and modeled by 
Falcke \& Markoff (2000), 
Yuan et al. 
(2003, 2009), Zhao et al. (2020), Brinkerink  
et al. (2021), 
Čemeljić 
et al. (2022), and 
Jiang et al. (2023), for example. As explained in detail in Section 3, 
the outflow method is independent of a detailed accretion disk or jet launching  
model. It is based on the premise that compact nuclear radio emission 
from sources that lie on the fundamental plane of black hole activity is 
associated with a jetted outflow, the outflow is powered at least in part by black hole spin energy extraction, and the mechanism is similar for all 
sources that lie on the fundamental plane.  

This approach has substantial empirical support. 
For example, the mapping of 
the fundamental plane of black hole activity 
to the fundamental line of black hole activity 
(Daly et al. 2018) 
leads 
to a relationship of the form $\rm{Log(L_{dKE}/L_{Edd}) = A ~Log(L_{bol}/L_{Edd}) +B}$.  As 
explained in section 5 of Daly et al. 
(2018), the most important parameter 
for understanding the physics of the sources, and defining an empirically-based model that describes the sources, is the parameter A. 
Consistent values of A (and B) are obtained for the four independent 
fundamental plane samples studied by  
Daly et al. (2018) (see Table 2 of that work), and the weighted mean value of A obtained for the ``combined sample" studied is $0.45 \pm 0.01$. 
Individual values and the combined value obtained with fundamental plane samples 
are in good agreement with those 
obtained with two completely independent 
methods including the ``cavity" method used by Merloni and Heinz (2007) for 
low-power extended radio sources 
and the strong shock method 
(e.g., O'Dea et al. 2009) 
used for 
high-power classical double (FRII) radio 
sources by Daly (2016). 
Merloni and Heinz (2007) obtain  
$\rm{Log(L_{dKE}/L_{Edd}) = (0.49 \pm 0.07) Log(L_{bol}/L_{Edd}) - (0.78 \pm 0.36)}$ where $\rm{L_{dKE}}$ is  obtained by dividing the P dV work required to inflate cavities and bubbles in hot X-ray emitting atmospheres of host galaxies and 
galaxy clusters by the buoyancy rise time, or the sound crossing time, or the refil time of the radio lobes (e.g., Birzan et al. 2004; Rafferty et al. 2006; Allen et al. 2006). 
For a sample of 97 classical double radio 
sources (i.e., FRII sources) including radio galaxies and radio loud quasars Daly (2016) obtains  
$\rm{Log(L_{dKE}/L_{Edd}) = (0.44 \pm 0.05) Log(L_{bol}/L_{Edd}) - (1.14 \pm 0.06)}$. This is consistent with the results of 
Merloni and Heinz (2007) and with results obtained with fundamental plane samples (Daly et al. 2018).

Thus, the beam power, bolometric luminosity, and black hole mass (i.e., Eddington luminosity) are 
fundamental physical variables that describe a black hole system 
for systems with collimated outflows, such as those described above. 
The general functional form that describes a spin-powered outflow 
(e.g., Blandford \& Znajek 1977; Moderski \& Sikora 1996; Meier 1999; 
Yuan \& Narayan 2014) is 
re-written in dimensionless separable functional form (see eq. 6 of Daly 2019), and 
terms between the two equations are 
identified, as explained in detail by 
Daly (2019). 
Identifying terms in these two equations allows the spin function, F, to be empirically determined for many types of sources including sources that lie on the fundamental plane of black hole activity such as Sgr A$^*$ and M87 
(Daly 2019). Clearly the method is independent of any particular accretion disk model, and of any particular spin-powered jet formation model. 
It does imply that the physical state of the accretion disk during the outflow 
is similar for all of the sources studied and 
can be parameterized as a function of the bolometric 
disk luminosity in Eddington units. It also implies that the 
spin energy extraction mechanism is similar for all of the sources, but it does not 
specify what that mechanism is. 
Thus, to solve for quantities in the context of a particular model, the model parameters should be re-cast in  dimensionless separable form, followed by the identification of terms between the empirical relationship and the theoretical relationship, as described in detail by Daly (2019). 

2). The collimated outflow may occur through the formation and 
release of unbound plasmoids, as 
discussed in detail by Jiang et al. (2023) 
(see also Comisso et al. 2017; 
Ripperda et al. 2022; McKinney 2006; 
Nakamura et al. 2018; Chatterjee et al. 2019; 
Borgogno et al. 2022
Nathanail et al. 2022). 
The formation and release of plasmoids 
is closely related to black hole spin and  
jet formation (Jiang et al. 2023) and  
fits in with global 
theoretical models of jet production such as 
the models of Blandford \& Znajek (1977), 
Punsly \& Coroniti (1990), Moderski \& Sikora (1996),  
Meier (1999), Punsly (2001), Nokhrina et al. (2019),  
Blandford \& Globus (2022), 
and Kino et al. (2022). Note that detailed analyses of the 
feasibility of the Blandford \& Znajek (1977) mechanism 
are presented and discussed by King \& Pringle (2021) and Komissarov (2022).  

The hotspot(s) observed in the vicinity of Sgr A* (GRAVITY Collaboration et al. 2018; 2020; Michail et al. 2021; 
Wielgus et al. 2022) 
have been interpreted as indicating the release 
of plasmoids (Nathanail et al. 2022b; Jiang et al. 2023).  
In the context of the outflow method described in section 3, this would 
contribute to the luminosity in directed 
kinetic energy, and would not contribute to the bolometric luminosity 
of the disk. That is, the outflow method separates the global properties of the disk 
from the properties of the outflow. In 
this model, 
the disk maintains the magnetic field that controls the extraction of spin energy 
from the black hole 
(as in the models of Blandford \& Znajek 1977;
Punsly \& Coroniti 1990; Moderski \& Sikora 1996; 
Meier 1999; Nokhrina et al. 2019; Kino et al. 2022; and 
Blandford \& Globus 2022, for example), and 
is parameterized in part by the bolometric 
luminosity of the disk. 

Thus, 
hotspots seen in the vicinity of Sgr A* 
that are taken to indicate the release of 
plasmoids (as discussed above) 
would be considered as contributions to the "jet power" 
in the context of the outflow method, 
and would not be considered as a contributing to the bolometric disk luminosity of the disk. 

3). As noted by Jiang et al. (2023), 
the frame-dragging effect of a rotating 
black hole amplifies the magnetic field and 
causes the field to accumulate close 
to the boundaries of the black hole 
(as expected theoretically based on the models mentioned above). 
This leads to the formation of plasmoids, many of which are unbound and hence flow 
away from the black hole region. 
This fits in nicely with the outflow method, 
which is a general empirically-motivated 
formulation that does not 
depend upon a specific jet production model 
or a specific accretion disk model, and is 
consistent with the 
Blandford \& Znajek (1977), 
Punsly \& Coroniti (1990), Moderski \& Sikora (1996),
Meier (1999), \& 
Blandford \& Globus (2022) models, for example.
Thus, both theoretical expectations and numerical simulations indicate that the 
power carried by unbound plasmoids is  strongly tied 
to how rapidly the black hole is spinning. 

As discussed by Jiang et al. (2023), the outflow associated with M87 is much more powerful in both  absolute and relative terms compared with that 
associated with Sgr A*. This is expected in the 
context of the outflow method. The application of the outflow method indicates that the spin angular momenta of 
M87 and Sgr A* are rather similar, $1.0 \pm 0.15$ and $0.93 \pm 0.15$, respectively 
(Daly 2019), with the value from Sgr A*  now 
updated to $0.90 \pm 0.06$ (see section 4). However, the  
spin functions, 
contributions of the rotational energy to the total black hole mass, and the 
spin mass-energy available for extraction relative to the total dynamical black hole mass differ both in relative and in 
absolute terms. Values for 
M87 are listed in Tables 4 and 5, and those for 
Sgr A* are listed in Tables 4, 5, 6, and 7. 
For example, the spin function of Sgr A* is about $\rm{F} = 0.62 \pm 0.10$ (see Table 6, preferred 
set I of this paper) while that for M87 is about 
$\rm{Log(F) = (0.13 \pm 0.19)}$ or 
$\rm{F} = 1.3 \pm 0.6$ (see Table 1 of Daly 2019); note that the spin function $\rm{F}$ is 
also equal to the ratio of the rotational 
mass to the irreducible mass, $\rm {F = (M_{rot}/M_{irr})}$, as shown by Daly (2022).
The spin mass-energy available for extraction relative to the total black hole mass is given by 
$[1-(F^2+1)^{-1/2}]$ (Daly 2022 eq. 10), leading 
to a weighted mean value of 
$\rm{(M_{spin}/M_{dyn})} \simeq 0.15 \pm 0.04$ for Sgr A* 
(see our "preferred" value given by Set I of Table 6, column 8), while that for M87 is $0.40 \pm 0.17$
(see Table 5). Thus, while about 15\% of the total dynamical mass of Sgr A* is 
available to be extracted, a 
much larger fraction, about 40\%, of the dynamical mass of M87 is available to be extracted. In absolute terms, this means that the mass-energy available for extraction 
for M87 is about $\rm{M_{spin} \simeq (2.6 \pm 1.2) \times 10^9 
M_{\odot}}$ assuming a black hole mass of about 
$(6.5 \pm 0.9) \times 10^9 \rm{M_{\odot}}$ 
(Event Horizon Telescope Collaboration 2019b), with 
the total uncertainty of the black hole mass estimated by adding the statistical and systematic uncertainties linearly.   
For Sgr A*, this value is about $\rm{M_{spin} \simeq (6.2 \pm 1.6) \times 10^5 ~M_{\odot}}$ 
(see column 9 of Table 6 for our preferred set I). 
Thus, both the fractional and absolute spin mass-energy available for extraction from M87 are substantially larger than that available for Sgr A*. 

Note that the high magnitude for the spin value for M87
discussed above is consistent with the analysis  presented by the Event Horizon Telescope 
Collaboration (2019a). However, a low 
spin value for M87 is obtained by Nokhrina et al. (2019). This could be interpreted in terms of the Moderski \& Sikora (1996) model, which explicitly includes the rotational speeds of the horizon and magnetic surfaces separately. 
For additional discussions see Blandford \& Globus (2022), 
Kino et al. (2022), and Hagen \& Done (2023), for example.  

4). One way to gauge the 
reliability of the outflow method is to check whether  consistent results are obtained with independently 
determined spin values, that is, those obtained with different methods. 
To date the outflow method has been applied to over 
700 supermassive black holes, and 102 measurements of four stellar-mass black holes, each of which is in an X-ray binary system (Daly 2019). 
Dimensionless spin angular momentum values 
obtained with the outflow method are compared with those obtained with independent methods for six supermassive black holes and 
two stellar-mass black holes by Daly (2019)  (see Table 1  and the discussion in section 5 of that work), and 
are compared 
by Azadi et al. (2023) for an additional 15  supermassive black holes. 

Spin values obtained for each of the 21 supermassive black holes for which a comparison was possible indicate excellent agreement between values obtained with two independent methods. For example, Azadi et al. (2023) apply the continuum fitting method to a sample of quasars with powerful outflows; 15 of the classical double radio sources in their sample overlap with those studied by Daly (2019) and in each case there is excellent agreement between the spin values obtained 
with independent methods. Six additional supermassive black holes studied by Daly (2019) 
had published dimensionless spin angular momentum values obtained with the X-ray reflection method and the comparison of independently determined spin values, listed in Table 1 of Daly (2019), indicate excellent agreement; the X-ray reflection values studied were published by Vasudevant et al. (2016), Patrick et al. (2012); and Walton et al. (2013). 

For stellar-mass black holes, a 
comparison of spin values obtained with the outflow method and an independent method was possible for two sources, GX 339-4 and AO6200 
(see Table 1 and the discussion in section 5 
of Daly 2019). For GX 339-4, 
the outflow method was applied to 
76 simultaneous radio and X-ray observations 
(Saikia et al. 2015) and a dimensionless spin 
value of $0.92 \pm 0.06$ is obtained for this black hole by Daly (2019) (see Table 1 of that work). This is in very good agreement with two independently determined values 
obtained with the X-ray reflection method of 
$0.94 \pm 0.02$ (Miller et al. 2009) and $0.95 ^{+0.03}_{-0.05}$ (Garcia et al. 2015). 

For the second stellar-mass system, AO6200, the dimensionless spin angular momentum value obtained 
by applying the outflow method to 
the data of Saikia et al. (2015), at which time 
the source was not in outburst, is $0.98 \pm 0.07$ (see Tables 1 and 2 of Daly 2019). The value obtained 
with the continuum fitting method 
using much earlier observations of the source during an unprecedented outburst 
is $0.12 \pm 0.19$ (Gou et al. 2010). Note 
that the continuum fitting method applied to the source 
is based on a particular accretion disk model of the source, and 
does not take into account the jetted outflow from the source.   
To address this discrepancy, Daly (2019) used X-ray and radio data obtained at the 
time of the outburst and  applied the outflow method 
to this data and obtained a value of $\rm{a_*} = 0.97 \pm 0.07$  
as discussed in detail in section 5 of Daly (2019). 
Thus, even though the disk luminosity differed by about six orders of magnitude between the two observations used to obtain the dimensionless spin angular momentum in the context of the outflow method, that method returned almost exactly the same value of $\rm{a_*}$ for that source. Daly (2019) interpreted this consistency as an indication that the results obtained with the outflow method are correct. 
One possible explanation of this discrepancy is that the continuum fitting method for this source was applied in the context of a particular accretion disk model, and this model may not provide an accurate description of the source during outburst. The outflow method is not based on a particular accretion disk model, or a particular jet production  model 
as explained above,  in point 1. 

Thus, a comparison of dimensionless spin angular momentum values obtained with the outflow method indicates very good agreement for all 21 supermassive black holes for which a comparison is possible; 
these comparisons included a comparison between results obtained with the continuum fitting method and the outflow method, and between the X-ray reflection method and the outflow method. For stellar mass black holes, excellent agreement was obtained for a comparison between the outflow method and the X-ray reflection 
method for GX 339-4.  

\section*{Acknowledgements}
We would like to thank the anonymous referee for very helpful comments and suggestions.  
It is a pleasure to thank Chetna Duggal for interesting discussions related to this work. 
The scientific results reported in this article are based on observations made by the Chandra X-ray Observatory and re-analyzed here. This research has made use of software provided by the Chandra X-ray Center (CXC) in the application package CIAO.
MD acknowledges the partial support of NASA-80NSSC21K1398 in this work.  

\section*{Data Availability}
The data underlying this article are available in the article or are available in the paper(s) referenced. 

\begin{table*}
\begin{minipage}{125mm}							
\caption{Data and Results for $\rm{Sgr~ A^*}$. }						
\label{Table4}								
\begin{tabular}{lllllcc}   
\hline\hline  
(1)&(2)&(3)&(4)&(5)&(6)&(7)\\	
&$	\rm{Log(L_R)}			$&$	\rm{Log(L_{dKE})}			$&$	\rm{Log(L_x)}			$&$	\rm{Log(L_{bol})}			$&\\
&5 GHz (erg/s)&(erg/s)&(2-10)~keV~(erg/s)&(erg/s)&~\rm{F}~\footnote{This is also equal to the rotational contribution to the total black hole mass  divided by the irreducible contribution to the total black hole mass (Daly 2022), $\rm{(M_{rot}/M_{irr}) = F}$, where the total black hole mass $\rm{M_{dyn}}$ is  
$\rm{M_{dyn} = (M_{rot}^2 +M_{irr}^2)^{1/2}}$, as discussed in the text.}			&{$a_*$}\\
\hline
C1&$	32.53	\pm	0.03	$&$	39.17	\pm	0.24	$&$	33.39	\pm	0.03	$&$	34.57	\pm	0.04	$&$	0.64	\pm	0.18	$&$	0.91	\pm	0.11	$\\
C2&$	32.45	\pm	0.03	$&$	39.11	\pm	0.24	$&$	33.48	\pm	0.03	$&$	34.66	\pm	0.04	$&$	0.58	\pm	0.16	$&$	0.86	\pm	0.12	$\\
C3&$	32.55	\pm	0.04	$&$	39.19	\pm	0.24	$&$	34.32	\pm	0.02	$&$	35.50	\pm	0.04	$&$	0.42	\pm	0.12	$&$	0.72	\pm	0.14	$\\
C4&$	32.43	\pm	0.01	$&$	39.10	\pm	0.24	$&$	33.16	\pm	0.05	$&$	34.34	\pm	0.06	$&$	0.66	\pm	0.18	$&$	0.92	\pm	0.10	$\\
\hline
B1&$	32.64	\pm	0.04	$&$	39.25	\pm	0.24	$&$	33.28	\pm	0.08	$&$	34.46	\pm	0.09	$&$	0.74	\pm	0.21	$&$	0.96	\pm	0.08	$\\
B2&$	32.52	\pm	0.05	$&$	39.16	\pm	0.24	$&$	33.48	\pm	0.03	$&$	34.66	\pm	0.04	$&$	0.61	\pm	0.17	$&$	0.89	\pm	0.11	$\\
B3&$	32.65	\pm	0.01	$&$	39.26	\pm	0.24	$&$	34.32	\pm	0.02	$&$	35.50	\pm	0.04	$&$	0.46	\pm	0.13	$&$	0.76	\pm	0.14	$\\
B4&$	32.51	\pm	0.01	$&$	39.16	\pm	0.24	$&$	33.41	\pm	0.04	$&$	34.59	\pm	0.05	$&$	0.63	\pm	0.17	$&$	0.90	\pm	0.11	$\\
\hline
M87~\footnote{The black hole spin function,  
F, and dimensionless spin angular momentum,  
$a_*$, obtained with the outflow method by Daly (2019) for M87 are included here for comparison. 
}&&&&&$1.3 \pm 0.6$&$1.0 \pm 0.2$\\
\hline
\hline
\end{tabular}									
\end{minipage}									
\end{table*}	

\begin{table*}								
\begin{minipage}{165mm}							
\caption{The Mass-Energy Components of $\rm{Sgr ~A}^*$ Obtained from the Spin Function, F. }						
\label{Table5}								
\begin{tabular}{llllllllll}   
\hline\hline  
(1)&(2)&(3)&(4)&(5)&(6)&(7)&(8)&(9)&(10)\\	
Data&$\rm{\underline{M_{rot}}}$&$\rm{M_{rot}}$\footnote{Obtained with a dynamical black hole mass 
of $(4.152 \pm 0.014) \times 10^6 ~\rm{M_{\odot}}$ for Sgr A* as discussed in the text.}&$\rm{\underline{M_{irr}}}$&$	\rm{M_{irr}}	$&$	\rm{\underline{M_{spin}}}	$&$	\rm{M_{spin}}$&$	\rm{\underline{M_{spin}}}	$&$\rm{\underline{M_{spin}}}	$&$\rm{\underline{\rm{E_{spin}}}}$\\
&$\rm{M_{dyn}}$&$(10^6 M_\odot)$&$\rm{M_{dyn}}$&$(10^6 M_\odot)$&$\rm{M_{dyn}}$&$(10^5 M_\odot)$&
$\rm{M_{rot}}$&$\rm{M_{irr}}$&$\rm{E_{spin,max}}$\\
\hline
C1	&$	0.54	\pm	0.11	$&$	2.2	\pm	0.4	$&$	0.84	\pm	0.07	$&$	3.5	\pm	0.3	$&$	0.16	\pm	0.07	$&$	6.6	\pm	2.8	$&$	0.29	\pm	0.07	$&$	0.19	\pm	0.10	$&$	0.45	\pm	0.23	$\\
C2	&$	0.50	\pm	0.10	$&$	2.1	\pm	0.4	$&$	0.87	\pm	0.06	$&$	3.6	\pm	0.3	$&$	0.13	\pm	0.06	$&$	5.5	\pm	2.5	$&$	0.27	\pm	0.06	$&$	0.15	\pm	0.08	$&$	0.37	\pm	0.19	$\\
C3	&$	0.39	\pm	0.09	$&$	1.6	\pm	0.4	$&$	0.92	\pm	0.04	$&$	3.8	\pm	0.2	$&$	0.08	\pm	0.04	$&$	3.3	\pm	1.6	$&$	0.20	\pm	0.05	$&$	0.09	\pm	0.05	$&$	0.21	\pm	0.11	$\\
C4	&$	0.55	\pm	0.11	$&$	2.3	\pm	0.4	$&$	0.83	\pm	0.07	$&$	3.5	\pm	0.3	$&$	0.17	\pm	0.07	$&$	6.9	\pm	2.9	$&$	0.30	\pm	0.07	$&$	0.20	\pm	0.10	$&$	0.48	\pm	0.24	$\\
\hline
B1	&$	0.59	\pm	0.11	$&$	2.5	\pm	0.5	$&$	0.80	\pm	0.08	$&$	3.3	\pm	0.3	$&$	0.20	\pm	0.08	$&$	8.1	\pm	3.3	$&$	0.33	\pm	0.07	$&$	0.24	\pm	0.12	$&$	0.59	\pm	0.30	$\\
B2	&$	0.52	\pm	0.11	$&$	2.2	\pm	0.4	$&$	0.85	\pm	0.06	$&$	3.5	\pm	0.3	$&$	0.15	\pm	0.06	$&$	6.1	\pm	2.7	$&$	0.28	\pm	0.07	$&$	0.17	\pm	0.09	$&$	0.41	\pm	0.21	$\\
B3	&$	0.42	\pm	0.10	$&$	1.7	\pm	0.4	$&$	0.91	\pm	0.04	$&$	3.8	\pm	0.2	$&$	0.09	\pm	0.04	$&$	3.8	\pm	1.8	$&$	0.22	\pm	0.05	$&$	0.10	\pm	0.05	$&$	0.24	\pm	0.13	$\\
B4	&$	0.53	\pm	0.11	$&$	2.2	\pm	0.4	$&$	0.85	\pm	0.07	$&$	3.5	\pm	0.3	$&$	0.15	\pm	0.07	$&$	6.3	\pm	2.8	$&$	0.29	\pm	0.07	$&$	0.18	\pm	0.09	$&$	0.44	\pm	0.22	$\\
\hline
&&$(10^9~M_{\odot})$&&$(10^9~M_{\odot})$&&$(10^9~M_{\odot})$\\
M87&$0.80 \pm 0.12$&$5.2 \pm 1.1$~\footnote{Obtained with a dynamical black hole mass of $(6.5 \pm 0.9) \times 10^9 M_{\odot}$ for M87 
as discussed in the text.}&$0.60 \pm 0.17$&$3.9 \pm 1.2$&
$0.40 \pm 0.17$&$2.6 \pm 1.2$&$0.50 \pm 0.13$&$0.68 \pm 0.47$&
$1.6 \pm 1.1$\\
\hline					
\hline							
\end{tabular}									
\end{minipage}									
\end{table*}

\begin{table*}								
\begin{minipage}{165mm}	
\caption{Weighted Mean Values of Quantities Obtained with Different Combinations of Data Sets for $\rm{Sgr~ A}^*$. }
\label{Table6}					
\begin{tabular}{lllllllll}   
\hline\hline 												
(1)&(2)&(3)&(4)&(5)&(6)&(7)&(8)&(9)\\		
Set&
&&						
$\rm{\underline{M_{rot}}}$&$\rm{M_{rot}}$&$\rm{\underline{M_{irr}}}$&$	\rm{M_{irr}}$&$	\rm{\underline{M_{spin}}}	$&$	\rm{M_{spin}}$\\
&$\rm{F_{WM}}$~\footnote{This is also equal to $\rm{(M_{rot}/M_{irr}})$; see Daly (2022).}
&$a_*$&
$\rm{M_{dyn}}$&$(10^6 M_\odot)$&$\rm{M_{dyn}}$&$(10^6 M_\odot)$&$\rm{M_{dyn}}$&$(10^5 M_\odot)$\\		
\hline
\hline
I~\footnote{Data Sets C1, C2, and C4; this is the ``preferred" combination of data sets, as discussed in the text. This includes the three non-flaring X-ray data sets obtained simultaneously with radio data sets, and each radio data set consist of a substantial number of observations.}
&$	0.62	\pm	0.10	$&$	0.90	\pm	0.06	$&$	0.53	\pm	0.06	$&$	2.2	\pm	0.3	$&$	0.85	\pm	0.04	$&$	3.5	\pm	0.2	$&$	0.15	\pm	0.04	$&$	6.2	\pm	1.6	$\\
II~\footnote{Data Sets C1, C2, C4, B1, and B4. This includes the three data sets that make up Set I, plus the two new X-ray non-flaring data sets; the radio data are obtained contemporaneously rather than simultaneously, and consist of one radio observation.}	&$	0.64	\pm	0.08	$&$	0.92	\pm	0.04	$&$	0.54	\pm	0.05	$&$	2.3	\pm	0.2	$&$	0.84	\pm	0.03	$&$	3.5	\pm	0.1	$&$	0.16	\pm	0.03	$&$	6.5	\pm	1.3	$\\
III~\footnote{Data Sets C1, C2, C3, and C4. This includes all four Chandra data sets obtained simultaneously with radio data sets, including the X-ray flaring data, and each radio data set consists of a substantial number of observations. }	&$	0.54	\pm	0.08	$&$	0.87	\pm	0.06	$&$	0.49	\pm	0.05	$&$	2.0	\pm	0.2	$&$	0.88	\pm	0.03	$&$	3.7	\pm	0.1	$&$	0.12	\pm	0.03	$&$	4.8	\pm	1.1	$\\
IV~\footnote{Data Sets C1, C2, C3, C4, B1, and B4. This includes the four data sets 
that make up Set III plus two X-ray non-flaring data sets, and thus includes 
all six X-ray data sets. It is the same as Set II plus the flaring X-ray data set 
combined with simultaneous radio data that consists of a substantial number of observations.}	&$	0.57	\pm	0.07	$&$	0.90	\pm	0.04	$&$	0.51	\pm	0.04	$&$	2.1	\pm	0.2	$&$	0.87	\pm	0.02	$&$	3.6	\pm	0.1	$&$	0.13	\pm	0.02	$&$	5.3	\pm	1.0	$\\
\hline						
\hline							
\end{tabular}									
\end{minipage}									
\end{table*}									

\begin{table*}								
\begin{minipage}{80mm}	
\caption{A Continuation of 
Weighted Mean Values of Quantities Obtained with Different Combinations of Data Sets for $\rm{Sgr A}^*$. }										
\label{Table7}					
\begin{tabular}{lllll}   
\hline\hline  								
(1)&(2)&(3)&(4)&(5)\\								
Set&
&$\rm{\underline{M_{spin}}}$&$\rm{\underline{M_{spin}}}$&$\rm{\underline{E_{spin}}}$\\							&$\rm{F_{WM}}$&$\rm{M_{rot}}$&$\rm{M_{irr}}$&$\rm{E_{spin,max}}$\\	
\hline											
I~\footnote{Data Sets C1, C2, and C4; this is the ``preferred" combination of data sets, as discussed in the text.}	&$	0.62	\pm	0.10	$&$	0.29	\pm	0.04	$&$	0.18	\pm	0.05	$&$	0.42	\pm	0.13	$\\
II~\footnote{Data Sets C1, C2, C4, B1, and B4.}	&$	0.64	\pm	0.08	$&$	0.29	\pm	0.03	$&$	0.19	\pm	0.04	$&$	0.45	\pm	0.10	$\\
III~\footnote{Data Sets C1, C2, C3, and C4.}	&$	0.54	\pm	0.08	$&$	0.26	\pm	0.03	$&$	0.12	\pm	0.03	$&$	0.30	\pm	0.08	$\\
IV~\footnote{Data Sets C1, C2, C3, C4, B1, and B4.}	&$	0.57	\pm	0.07	$&$	0.27	\pm	0.03	$&$	0.14	\pm	0.03	$&$	0.33	\pm	0.08	$\\
\hline																	
\hline																	
\end{tabular}																	
\end{minipage}																	
\end{table*}



\bibliographystyle{mnras}
\bibliography{example} 







\bsp	
\label{lastpage}
\end{document}